# Simulating high-temperature superconductivity in moiré WSe$_2$

Yiyu Xia[1*], Zhongdong Han[2*], Jiacheng Zhu[1], Yichi Zhang[2], Patrick Knüppel[2], Kenji Watanabe[3], Takashi Taniguchi[3], Kin Fai Mak[1,2,4,5**], and Jie Shan[1,2,4,5**]

[1]School of Applied and Engineering Physics, Cornell University, Ithaca, NY, USA
[2]Laboratory of Atomic and Solid State Physics, Cornell University, Ithaca, NY, USA
[3]National Institute for Materials Science, Tsukuba, Japan
[4]Kavli Institute at Cornell for Nanoscale Science, Ithaca, NY, USA
[5]Max Planck Institute for the Structure and Dynamics of Matter, Hamburg, Germany

*These authors contributed equally
**Email: kin-fai.mak@mpsd.mpg.de; jie.shan@mpsd.mpg.de

**The emergence of high transition temperature ($T_c$) superconductivity in strongly correlated materials remains a major unsolved problem in physics. High-$T_c$ materials, such as cuprates, are generally complex and not easily tunable, making theoretical modelling difficult. Although the Hubbard model--a simple theoretical model of interacting electrons on a lattice[1]--is believed to capture the essential physics of high-$T_c$ materials[2-8], obtaining accurate solutions of the model, especially in the relevant regime of moderate correlation, is challenging[9,10]. The recent demonstration of robust superconductivity in moiré WSe$_2$ (Ref. [11,12]), whose low-energy electronic bands can be described by the Hubbard model and are highly tunable[13-15], presents a new platform for tackling the high-$T_c$ problem. Here, we tune moiré WSe$_2$ bilayers to the moderate correlation regime through the twist angle and map the phase diagram around one hole per moiré unit cell ($\nu = 1$) by electrostatic gating and electrical transport and magneto-optical measurements. We observe a range of high-$T_c$ phenomenology, including an antiferromagnetic insulator at $\nu = 1$, superconducting domes upon electron and hole doping, and unusual metallic states at elevated temperatures including strange metallicity[16-18]. The highest $T_c$ occurs adjacent to the Mott transition[5,19], reaching about 6% of the effective Fermi temperature. Our results establish a new material system based on transition metal dichalcogenide (TMD) moiré superlattices that can be used to study high-$T_c$ superconductivity in a highly controllable manner and beyond.**

The Hubbard model[1], describing electrons hopping on a lattice with a hopping amplitude $t$ between neighboring sites and an on-site electron-electron repulsion $U$, provides a simplified representation of high-$T_c$ materials[2-8]. The electron hopping leads to a finite bandwidth $W$ ($= 8t$ and $9t$, respectively, for square and triangular lattices). In cuprates, $W$ and $U$ are comparable[5,8], indicating that these materials are in the moderate correlation regime and are near a Mott transition from an insulator to a metal[5,19]. The idea of doping a Mott insulator for high-$T_c$ superconductivity has been extensively studied for decades[2-8]. Although the nature of the superconducting state itself has been qualitatively understood, a full understanding of the rich phase diagram remains elusive because accurately solving the Hubbard model in the moderate correlation regime, where different orders intricately compete in the ground state, is difficult. Simulating high-$T_c$ phenomenology in a controlled quantum system[20,21] may shed new light on how superconductivity emerges and possibly pave the way for designing new high-$T_c$ materials.

TMD moiré heterobilayers, such as $WSe_2/WS_2$, are established quantum simulators of the two-dimensional (2D) triangular lattice Hubbard model[22-26]. For small twist angles, the moiré period (consequently $U/W$) is largely determined by the lattice mismatch for heterobilayers. Studies so far have focused on the strong correlation limit with large $U/W$ and superconductivity has yet been realized. The recent demonstration of superconductivity in twisted $WSe_2$ ($tWSe_2$) homobilayers[11,12], for which $U/W$ can be readily tuned by the sample twist angle[27-30], provides the possibility of exploring the superconducting phase diagram in the moderate correlation regime[31-35]. Here, we report the phase diagram of $tWSe_2$, focusing on 4.6° $tWSe_2$, for which $W$ and $U$ are comparable. The transport characteristics, supplemented by magnetic susceptibility, reveal rich high-$T_c$ phenomenology, including an antiferromagnetic (AF) insulator, superconducting domes, strange metals, and possibly also a 'pseudogap' phenomenon. The phase diagrams for both electron and hole doping of the AF insulator and their continuous evolution as the system undergoes a band-structure-tuned Mott transition can be obtained using a single dual-gated device.

**Twist angle effects**
Monolayer $WSe_2$ is a triangular lattice semiconductor with the valence band maxima at the K and K' points of the hexagonal Brillouin zone (BZ)[26]. A triangular moiré lattice with period $a_M = \frac{a}{\sqrt{2(1-\cos\theta)}}$ is formed when two layers are stacked with a relative twist angle $\theta$, where $a = 0.33$nm is the monolayer lattice constant. Figure 1b and 1c show the band structure and density of states (DOS) as a function of hole filling factor ($\nu$) and vertical electric field ($E$) for 4.6° $tWSe_2$ from the continuum model calculations (Methods). The K-valley states of the top and bottom layers fold, respectively, onto the $\kappa$ and $\kappa'$ valleys of the moiré BZ, which are swapped for the K'-valley states[13-15,36,37]. The bands from the two layers hybridize and generate a saddle point where they intersect. For small twist angle, the hybridized bands remain spin degenerate because of spin-valley locking in monolayer TMDs[38] (top panel). Application of a finite $E$-field places the two layers at different potentials and lifts the degeneracy[14] (bottom panel). As $E$ increases, the van Hove singularity (vHS) with a diverging DOS shifts continuously from $\nu < 1$ to $\nu > 1$. Sufficiently large fields can eventually polarize the holes to one layer (Fig. 1c). Theoretical studies showed that the topmost moiré valence band, if nontopological, can be described by the triangular lattice Hubbard model with an additional $E$-dependent spin-orbit coupling (SOC) term[13-15]. This applies to samples with twist angle larger than about 4° (Ref. [39]).

Figure 1d demonstrates the twist-angle effects on DC transport in $tWSe_2$. The sample resistance $R$ was measured at temperature $T = 1.6$K as a function of $E$ and $\nu$ using the dual-gated device structure shown in Fig. 1a (see Methods for details on the device fabrication and characterizations and Extended Data Fig. 1 for a device image). The resistance map for different twist angles qualitatively agrees with the DOS map in Fig. 1c, including a layer-hybridized region centered at $E = 0$ and a vHS with enhanced resistance. Not captured by the band theory are the insulating states in the layer-hybridized region at factional fillings of the first moiré band, $\nu = 1, \frac{1}{3}, \frac{1}{4}$, which manifest strong electron correlations. As twist angle increases from 2.7° to 4.9°, the correlated insulators melt and become indiscernible at 4.9°. This is expected because the correlation effects weaken with decreasing moiré period (increasing twist angle). The $E$-field dependence of the $\nu = 1$ state is further enriched by the presence of the vHS, which enhances the correlation effects[11,12,14,27,40].

Superconductivity was reported in two types of tWSe$_2$ devices. The first is around twist angle 3.6°, where the correlated insulators dominate the layer-hybridized region, and the $E$-field induces a transition from a Mott insulator to a superconductor around $\nu = 1$ near zero $E$-field[11]. The second is around twist angle 5°, where there are no correlated insulators, and superconductivity is observed along the vHS over a range of $\nu \approx 1 - 1.2$ (Ref. [12]). Both cases were examined theoretically[41-56]. These earlier results suggest that superconductivity exists in tWSe$_2$ over a range of twist angle. Although the superconducting state itself may be of the same nature, samples with intermediate twist angle and moderate correlation may allow for studies of the intricate phase diagram as in high-$T_c$ materials[3,5-8].

**Superconductor and AF insulator**
We focus on tWSe$_2$ around 4.6°, where the correlated insulator at $\nu = 1$ is present only near the region intersected by the vHS. Figure 2a shows resistivity $\rho_{xx}$ as a function of $E$ and $\nu$ for device 1 at 50mK. Superconductivity emerges near the insulator. Figure 2b is a close-up view of the dashed box in Fig. 2a. Figure 2c displays Hall resistivity $\rho_{xy}$ of the same region measured under an out-of-plane magnetic field $B = 0.3$T. The Hall resistivity for the insulator (shaded in grey) cannot be accessed due to its large $\rho_{xx}$. The superconductor wraps around the insulator with an intermediate metallic phase. The metal exhibits a significantly enhanced $\rho_{xy}$, which is consistent with electron conduction for $\nu < 1$ and hole conduction for $\nu > 1$. More discussions on this phase will follow in the next section.

The superconductor becomes more robust towards its boundary with the insulator (Extended Data Fig. 3). The highest $T_c$, whose location is denoted by an empty circle in Fig. 2b, is about 400mK. Detailed characterizations of the state are shown in Extended Data Fig. 4. At 50mK, the critical $B$-field is about 0.15T. The coherence length is $\xi \approx$ 34nm from the Ginzburg-Landau analysis of the $T$-dependence of the critical field. Given the moiré period $a_M \approx 4.1$nm and the electron mean free path $l \gtrsim 400$nm (estimated from the normal-state $\rho_{xx}$), the Cooper pairs are tightly bound, and the superconductor is in the clean limit. The superconductor from other parts of the phase space has similar properties.

We examine the nature of the insulator at $\nu = 1$ by probing the magnetic susceptibility $\chi_{MCD}$, which is determined as the weak-field slope of the magnetic circular dichroism[24] (Methods). Figure 2d shows the reflection contrast (RC) of the tWSe$_2$ moiré exciton as a function of $E$ and $\nu$ at 1.6K for device 2 (4.5°), which is without the contact and split metal gates for optical access. The insulator is identified by an enhanced RC for $E$ between 70 and 120 mV/nm. Figure 2e shows the $T$-dependence of $\chi_{MCD}$ at $\nu = 1$ for selected $E$- fields. The curves are vertically displaced for clarity. The susceptibility with a cusp observed at $E = -102, -112, -122$ mV/nm is characteristic of an antiferromagnet. The cusp marks the Néel temperature $T_N$ ($\approx 8$ K), at which the antiferromagnet orders. The susceptibility also follows the Curie-Weiss law for $T > T_N$ with a negative Curie-Weiss temperature. As $E$ moves the system closer to the Mott transition, $T_N$ continuously decreases. Beyond the transition ($E = -133$ mV/nm), $\chi_{\text{MCD}}$ diverges at low temperatures and there is no longer long-range magnetic order. A similar trend is observed in Fig. 2f for doping away from $\nu = 1$ at a fixed $E$-field. More susceptibility data can be found in Extended Data Fig. 5. These results indicate that the insulator at $\nu = 1$ is an AF insulator and the AF order persists, albeit weakened, when small doping is introduced. Although the exact value of $T_N$ and the $E$-field range for the insulator depend on the sample twist angle and other details, the AF insulating phase is generic for tWSe$_2$ with moderate correlation at $\nu = 1$. Theoretical studies showed that the insulator can be described by the 2D anisotropic XXZ model with an $E$-field-dependent Dzyaloshinskii–Moriya interaction term, and the spin anisotropy favors a 120-degree Néel order[13-15,40].

**Phase diagram**

We map the phase diagram of 4.6° tWSe$_2$ for four representative $E$-fields that cross the Mott transition at $\nu = 1$ as denoted by the arrows in Fig. 2b. Figure 3 displays $\rho_{xx}$ at $B = 0$T as a function of temperature and filling (top row), the phase diagrams (middle), and the Hall density as a function of filling at 50mK (bottom). Detailed line cuts and analysis are shown in Fig. 4 and Extended Data Fig. 6-9. A temperature-$E$-field phase diagram is included in Extended Data Fig. 10 for $\nu = 1$. We first focus on the case of $E = -103$ mV/nm. The $\rho_{xx}$ map (Fig. 3a) shows a prominent insulator at $\nu = 1$ and a superconducting dome on each side of the insulator with a minimum doping about 0.03 from the insulator. Superconductivity below $\nu = 1$ is more robust. The $\rho_{xy}$ map (Extended Data Fig. 6) shows a significant enhancement at low temperatures for fillings between the insulator and the superconductor.

Figure 4a-h are the linecuts of Fig. 3a at selected fillings. At $\nu = 1$, $\rho_{xx}$ decreases with decreasing temperature till about 3.5K, below which it increases sharply by nearly three orders of magnitude (Fig. 4a). Together with the magnetic susceptibility result above, this supports a transition from a paramagnetic metal to an AF insulator at $T_N \approx 3.5$K. When small doping is introduced (Fig. 4b,c), the sample becomes metallic: $\rho_{xx}$ decreases with temperature except a sharp peak or increase around 2.5 K. The feature is accompanied by an abrupt jump of the Hall density $n_H$ by about a moiré density $n_M$. The Hall density approaches $0.02 n_M$ and $-0.02 n_M$ (denoted by dashed lines) at low temperatures for $\nu = 0.98$ and 1.02, respectively. These results demonstrate a phase transition between two distinct metallic states with an abrupt Fermi surface reconstruction. The low-temperature phase has a small Fermi surface with carrier density $(1 - \nu) n_M$ and is antiferromagnetically ordered (Fig. 2f). It is a doped AF insulator, that is, the Mott gap survives upon doping[5]. We determine $T_N$ as the onset temperature for the $n_H$ jump, at which $\rho_{xx}$ is a local minimum. The Néel temperature continuously decreases with doping away from the AF insulator and vanishes at $\nu \approx 0.96$ and 1.04, which are near the peaks of the superconducting domes (Fig. 3e).

For a wide doping range of $0.88 < \nu < 0.96$, which includes the entire superconducting dome, $\rho_{xx}$ follows a $T$-linear dependence above a temperature $T'$ and a sub-linear dependence below $T'$ (Fig. 4d). We define the crossover temperature scale $T'$ as the temperature at which $\rho_{xx}$ deviates from the $T$-linear dependence by 10%. This temperature has a dome-like dependence on filling and vanishes near $\nu = 0.88$ (Fig. 3e).

For $\nu < 0.88$, a Fermi liquid behavior emerges at low temperatures with $\rho_{xx} = \rho_0 + AT^2$ (Fig. 4h), where $\rho_0$ is the residual resistivity and $A$ is often used to track the quasiparticle effective mass $m^*$. The resistivity deviates from the $T^2$-dependence above the coherence temperature $T_{coh}$, which is defined as the temperature for 10% deviation. The coherence temperature vanishes near $\nu = 0.88$ and increases with decreasing $\nu$ (Fig. 3e). Accordingly, $A$ diverges near the same filling (Fig. 3i).

Figure 4i compares the $T$-dependence of $\rho_{xx}$ for different metallic states around $\nu = 0.88$. $T$-linear resistivity over the entire temperature range of this study (50mK-20K) is observed only at $\nu = 0.88$, which separates two metals with vanishing $T_{coh}$ and $T'$. In addition, $\rho_{xx}/\rho_{xy}$, which relates to the Hall angle, scales quadratically with temperature (Fig. 4f). Such behaviors are like those of 'strange metals' found in strongly correlated materials near the quantum critical points[16-18,57-62]. Further, the dependence has a slope around 100Ω/K and a residual resistivity below 100Ω. The scattering time from a Drude analysis using the measured $m^*$ and $n_M$ (Extended Data Fig. 2) is close to the Planckian time (Methods). The small residual resistivity shows that the strange metal is in the clean limit.

The phase diagram for hole doping ($\nu > 1$) is similar except that the temperature scale $T'$ and the metallic state associated with it cannot be identified (Fig. 4c,e and Extended Data Fig. 6). $T$-linear resistivity for the entire temperate range (above $T_c$) is also observed near $\nu = 1.05$, where $T_{coh}$ and $T_N$ vanish (Fig. 4e), but the accompanying $\rho_{xx}/\rho_{xy}$ does not follow the simple $T^2$-dependence as for $\nu = 0.88$.

The phase diagram (Fig. 3e) bears a remarkable resemblance to, and, at the same time, subtle differences from that of the high-$T_c$ cuprates[3-8]. Doping an AF insulator with a small number of electrons or holes yields an AF metal with a small Fermi surface. Near the optimal doping for superconductivity, the Mott gap collapses, manifesting a Hall density jump[63] by about $n_M$ (Fig. 3i). Before the Mott gap collapses, $T_c$ is about 6% of the effective Fermi temperature $T_F$ (Methods), which is on par with most unconventional superconductors[64]. The Hall density jump and the suppressed $\chi_{MCD}$ for $T < T_N$ suggest the AF metal regions immediately next to $\nu = 1$ are similar to the 'pseudogap' phase in cuprates[5,65], but future studies showing a spectral gap in these regions are required to verify this assignment. While the phase diagram for $\nu > 1$ is similar to the cuprates, that for $\nu < 1$ is less so. In particular, there is an additional $T'$ dome showing unconventional $T$-dependence of $\rho_{xx}$, whose physical significance remains to be understood; the strange metal at $\nu = 0.88$ is also outside the superconducting dome. These differences from the cuprates may originate from the different lattice symmetry (triangular versus square), which deserves further investigations.

**Phase diagram through the Mott transition**

Finally, we examine the evolution of the phase diagram through the $E$-field-tuned Mott transition at $\nu = 1$ with critical field $E_c \approx -92$ mV/nm. As $E$ varies from -103mV/nm (Fig. 3a,e,i) to -99mV/nm (Fig. 3b,f,j) to -96mV/nm (Fig. 3c,g,k), the system moves closer to the transition from the insulator side. The phase diagram remains qualitatively unchanged but $T_N$ decreases; the doped AF insulator shrinks; the superconducting domes expand and $T_c$ increases. Also observed are the Hall density jumps by about $n_M$ near the optimal doping levels and Fermi liquids at sufficiently large doping levels (whose onset is marked by a divergent $A$). However, across the Mott transition at $E = -87$ mV/nm (Fig. 3d,h,l), the AF insulator disappears and only one superconducting dome remains. The Hall density jump at optimal doping becomes about $2n_M$ which is associated with the vHS in the band structure. An unusual feature is that $T$-linear resistivity persists over a wide filling range of $0.86 < \nu < 1.01$ (Fig. 4j) and Fermi liquids arise outside this range. Similar extended filling range for $T$-linear resistivity is also observed at $E = -96$ mV/nm (Extended Data Fig. 8). The results suggest that strange metallicity could occur over a range of tuning parameter instead of a quantum critical point[65-67] (see Extended Data Fig. 9 and 10 for additional results and analysis). The observation highlights the complexity of the phase diagram near the Mott transition and calls for experiments beyond DC transport. We summarize the interplay between the AF order and superconductivity in Extended Data Fig. 11. As $E$ approaches $E_c$ from the insulator side, $T_N$ decreases and the optimal $T_c$ increases monotonically; $T_c/T_N$ increases from 4% (at -110mV/nm) to 18% (at -96mV/nm) and is expected to diverge at $E_c$.

**Conclusion and outlook**
Tuning tWSe$_2$ to the moderate correlation regime through its twist angle, we observe a phase diagram around $\nu = 1$ that resembles the iconic phase diagram of the high-$T_c$ cuprates[3-8]. Superconductivity is observed only near the gate-tuned Mott transition, consistent with numerical studies of the Hubbard model[68]. The observations suggest that pure phonon-mediated Cooper pairing is unlikely. They promote the strong correlation viewpoint for high-$T_c$ superconductivity. Additional experimental probes are required to further understand the complex phase diagram. The platform based on the TMD moiré superlattices could provide new perspectives on the high-$T_c$ problem. Topological properties that are intrinsic among these materials also open the possibility of realizing quantum phenomena that are absent in known high-$T_c$ materials and junctions.

**Methods**
**Device geometry and fabrication.**
The tWSe$_2$ field-effect devices use contact gates and split gates (Fig. 1a and Extended Data Fig. 1) to achieve low contact resistances (4kΩ typically) and to turn off unwanted parallel conduction channels, respectively. The channel of the Hall-bar devices is defined by gates, including the top, bottom, contact and split gates. Details of the device geometry and device fabrication have been described in Ref. [11]. In brief, 2D flakes, including few-layer graphite, hexagonal boron nitride (hBN) and monolayer WSe$_2$, were mechanically exfoliated from bulk crystals onto silicon substrates. Large WSe$_2$ monolayers were cut into half by an atomic force microscope tip. These flakes were sequentially picked up using a polycarbonate stamp. From top to bottom, the heterostructures consist of a hBN capping layer, a graphite and hBN (3.4nm) top gate, the first and second half of a WSe$_2$ monolayer with a small twist angle, and a hBN (12.0nm) and graphite bottom gate. The heterostructures were released onto silicon substrates with pre-patterned titanium/platinum (5nm/25nm) electrodes. Titanium/palladium (5nm/35nm) contact gates and split gates were deposited using the standard electron-beam lithography and evaporation techniques.

**Electrical measurements.**
The electrical transport measurements were performed in a dilution refrigerator (Bluefors LD250) equipped with a 12T superconducting magnet. Silver-epoxy filters (Basel Precision Instruments MFT25) and additional RC (resistor–capacitor) filters were installed on the mixing chamber plate to ensure sufficient thermalization and efficient filtering of high-frequency radiation, achieving a base electron temperature of less than 100mK. The standard low-frequency (5.777Hz) lock-in technique was employed to measure both the longitudinal and transverse resistances. The excitation current was kept at 1 nA to minimize sample heating. Both the voltage drop at the probe electrodes and the source-drain current were recorded to determine the resistance. The sample resistivity, $\rho_{xx} = R_{xx}/(L/W)$, was calculated from the longitudinal resistance $R_{xx}$, where $L$ ($\approx 1.6\mu m$) is the center-to-center distance between the voltage probes and $W$ ($\approx 1.3\mu m$) is the channel width. The Hall resistivity $\rho_{xy}$ is equal to the anti-symmetrized transverse resistance $R_{xy}$ under a positive and negative $B$-field.

**Magneto-optical measurements.**
Magneto-optical measurements were performed in a closed-cycle cryostat (Attocube, Attodry 2100) at $B$-fields up to 9T and temperatures down to 1.6K. Details of the measurements have been reported in Ref. [30]. In short, a linear polarizer and a quarter-wave plate were used to generate circularly polarized light ($\sigma^+$ and $\sigma^-$) from a light-emitting diode. The polarized light beam was focused onto the sample by a low-temperature objective lens (Attocube, numerical aperture 0.8) with a spot size about 1μm. The excitation intensity was kept below 50nW/μm² to minimize sample heating. No measurable changes in MCD were recorded upon further reducing the excitation power by an order of magnitude. The reflected signal was collected by the same objective and directed to a spectrometer equipped with a liquid-nitrogen-cooled charge-coupled device (CCD) for spectral acquisition. The MCD spectrum is defined as $\frac{(I^- - I^+)}{(I^- + I^+)}$, where $I^-$ and $I^+$ are the reflection spectra for the $\sigma^-$ and $\sigma^+$ incident light, respectively. To obtain the data shown in the figures, we integrated the MCD signal over a narrow spectral window covering the moiré exciton resonance (735–740 nm) of tWSe$_2$. The magnetic susceptibility was extracted by a linear fit to the $B$-field dependence of the MCD signal near zero $B$-field. The reflection contrast spectrum is defined as $(I - I_0)/I_0$, in which $I$ is the reflection spectrum at arbitrary doping densities and $I_0$ is the reference spectrum at a high doping density, where no distinct excitonic resonances can be observed. To obtain the reflection contrast map in Fig. 2d, we used the mean value of $(I - I_0)/I_0$ over the spectral window (736–741 nm) that covers the moiré exciton resonance.

**Band structure calculations.**

The low-energy electronic band structure of small twist angle tWSe$_2$ was calculated using the continuum model, as described in Refs. [36,37]. Monolayer WSe$_2$ is a direct-gap semiconductor with the bandgap located at the two inequivalent corners of the hexagonal Brillouin zone (BZ), K and K′. In monolayer WSe$_2$, the valley degree of freedom of the electrons is locked to its spin due to the broken inversion symmetry and large spin-orbit coupling[38]. The K and K′ states, carrying opposite spin polarizations, are related by time-reversal symmetry. In tWSe$_2$, the valley pocket K$_t$ and K$_b$, originated from the top and bottom monolayers, respectively, are slightly displaced in momentum and define the corners of the moiré BZ for each spin flavor. The low-energy physics of hole-doped tWSe$_2$ is captured by a two-band **k · p** model within an effective mass description.

The effective moiré Hamiltonian for the K-valley (spin-up) states can be written as

$$H_\uparrow = \begin{pmatrix} -\frac{\hbar^2(\boldsymbol{k}-\boldsymbol{\kappa}_+)^2}{2m^*} + \Delta_t(\boldsymbol{r}) & \Delta_T(\boldsymbol{r}) \\ \Delta_T^\dagger(\boldsymbol{r}) & -\frac{\hbar^2(\boldsymbol{k}-\boldsymbol{\kappa}_-)^2}{2m^*} + \Delta_b(\boldsymbol{r}) \end{pmatrix}. \quad (1)$$

The Hamiltonian $H_\downarrow$ for the K′-valley (spin-down) states can be obtained by a time-reversal transformation of $H_\uparrow$. In Eq. (1), $\boldsymbol{k}$ and $\boldsymbol{r}$ denote, respectively, the wave vector and position vector of the holes; $\boldsymbol{\kappa}_\pm$ represent the corners of the moiré BZ; $\hbar$ is the reduced Planck constant; $m^*$ is the hole effective mass (0.45 $m_0$ for the highest valence band of monolayer WSe$_2$)[69]. ($m_0$ is the free electron mass.) The holes are subjected to a periodic pseudomagnetic field, $\Delta(\boldsymbol{r}) = (\text{Re}\Delta_T^\dagger, \text{Im}\Delta_T^\dagger, \frac{\Delta_t-\Delta_b}{2})$, with the moiré period. Within the lowest harmonic approximation, $\Delta_T(\boldsymbol{r}) = w(1 + e^{-i\boldsymbol{g}_2\cdot\boldsymbol{r}} + e^{-i\boldsymbol{g}_3\cdot\boldsymbol{r}})$ is the interlayer tunneling amplitude, and $\Delta_{t,b}(\boldsymbol{r}) = \pm\frac{V_z}{2} + 2V\sum_{j=1,3,5}\cos(\boldsymbol{g}_j\cdot\boldsymbol{r}\pm\psi)$ are the intralayer moiré potential for the top and bottom layers, respectively. Here $V_z$ is the sublattice potential difference; $\boldsymbol{g}_{j=1,2,3}$ are the reciprocal lattice vectors obtained by rotating $\boldsymbol{g}_1 = (\frac{4\pi}{\sqrt{3}a_M}, 0)$ counterclockwise by an angle of $(j-1)\pi/3$. The parameters $(V, \psi, w) = (13.6\text{meV}, 49.1°, 10.0\text{meV})$ are adopted from a recent scanning tunneling microscopy study on tWSe$_2$ (Ref. [39]).

The band structure was obtained by diagonalizing the Hamiltonian, using which the DOS was computed. For comparison with experiment, the sublattice potential difference is converted to an out-of-plane $E$-field using the relation, $E = V_z/(\frac{\varepsilon_{hBN}}{\varepsilon_{TMD}}et)$, where the dipole moment $\frac{\varepsilon_{hBN}}{\varepsilon_{TMD}}et \approx 0.26e$ nm was independently determined from the anti-crossing feature of the layer-hybridized moiré excitons, following Ref. [70,71]. The value is consistent with the accepted out-of-plane dielectric constants of hBN and WSe$_2$ ($\varepsilon_{hBN} \approx 3$ and $\varepsilon_{TMD} \approx 8$) and the interlayer separation in WSe$_2$ ($t \approx 0.7$nm).

**Estimate of $T_c/T_F$ at $E = -103$ mV/nm.**
The superconducting dome at fillings below $\nu = 1$ overlaps with the doped AF phase at low temperatures (see Fig. 3a,e and Extended Data Fig. 6). We can estimate the ratio $\frac{T_c}{T_F}$ using $T_F \approx \frac{\hbar^2 \pi |1-\nu| n_M}{m^* k_B}$ in this overlapped region. Here $k_B$ denotes the Boltzmann constant, and $n_M \approx 6.7 \times 10^{12}\text{cm}^{-2}$ is the moiré density. The Fermi temperature $T_F \approx 5K$ is that for a small Fermi surface (with density $|1-\nu|n_M \approx 0.04 n_M$). Using an effective mass $m^* \approx 1.3 m_0$ from $T$-dependent quantum oscillation measurements (Extended Data Fig. 2), we estimate $\frac{T_c}{T_F} \approx \frac{0.3K}{5K} \approx 6\%$. The ratio is comparable to $\frac{T_c}{T_N} \approx \frac{0.3K}{3.5K} \approx 8\%$ (see Extended Data Fig. 11). These estimates show the emergence of high-$T_c$ superconductivity relative to $T_F$ and $T_N$.

**Possible Planckian dissipation in strange metals.**

The strange metals in our study show a $T$-linear dependence over a wide temperature range. The $T$-linear dependence has a typical slope $\frac{\Delta \rho_{xx}}{\Delta T} \approx 100\Omega/K$ and a residual resistivity below $100\Omega$. We can estimate the scattering time $\tau$ from the slope assuming Boltzmann transport and compare it to the Planckian timescale to obtain the proportionality constant[18], $\alpha \equiv \frac{h}{\tau k_B T} \approx \frac{h^2 n_M}{m^* k_B} \frac{e^2}{h} \frac{\Delta \rho_{xx}}{\Delta T} \approx 1$. Here $h$ and $e$ denote the Planck constant and electron charge, respectively; the effective mass $m^* \approx m_0$ is determined by $T$-dependent quantum oscillation studies (Extended Data Fig. 2); and the carrier density is approximated by $n_M \approx 6.7 \times 10^{12}\text{cm}^{-2}$ since we examine fillings near $\nu = 1$. Combined with the small residual resistivity $\rho_0$, the observed strange metal is in the clean limit and possibly exhibits Planckian dissipation.

**Acknowledgements**
We thank Patrick Lee, Andrew Millis, Liang Fu, Piers Coleman, Michael Lawler, Brad Ramshaw, Debanjan Chowdhury, Senthil Todadri, Qimiao Si, Ashvin Vishwanath, Andrei Bernevig, Jedediah Pixley, Allan MacDonald and Philip Phillips for many helpful discussions.

# Figures

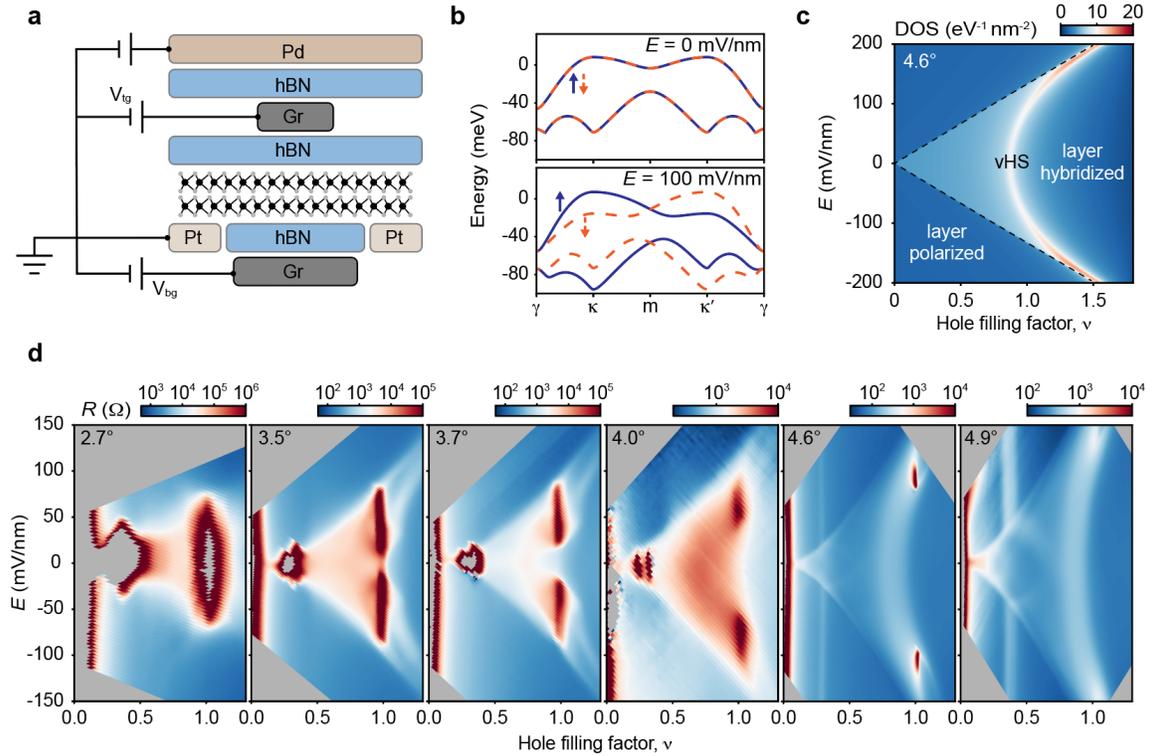

**Figure 1 | Twist angle effects. a,** Schematic side view of dual-gated transport devices. The tWSe$_2$ sample is contacted by platinum (Pt) electrodes and controlled by two voltages ($V_{tg}$ and $V_{bg}$) applied on the hexagonal boron nitride (hBN)/graphite (Gr) gates. The palladium (Pd) contact and split gates (details not shown) are used to turn on the Pt contacts and turn off the parallel channels, respectively. **b,** Topmost moiré valence bands for the K-valley state (blue solid line with spin up) and K'-valley state (orange dashed line with spin down) at $E = $ 0mV/nm (upper panel) and 100mV/nm (lower panel). **c,** Electronic density of states (DOS) versus $\nu$ and $E$. The van Hove singularity (vHS) with high DOS shifts towards higher $\nu$ with increasing $E$. Dashed lines denote the boundary between the layer-hybridized and layer-polarized regions. Results in **b,c** are from the continuum model calculations for 4.6° tWSe$_2$. **d,** Longitudinal resistance $R$ of tWSe$_2$ versus $\nu$ and $E$ at $T = $ 1.6K and $B = $ 0T for twist angle ranging from 2.7º to 4.9º. Correlated insulators are observed at $\nu = 1, \frac{1}{3}, \frac{1}{4}$ in small twist angle samples and absent in the 4.9° sample. The correlation effects weaken with increasing twist angle.

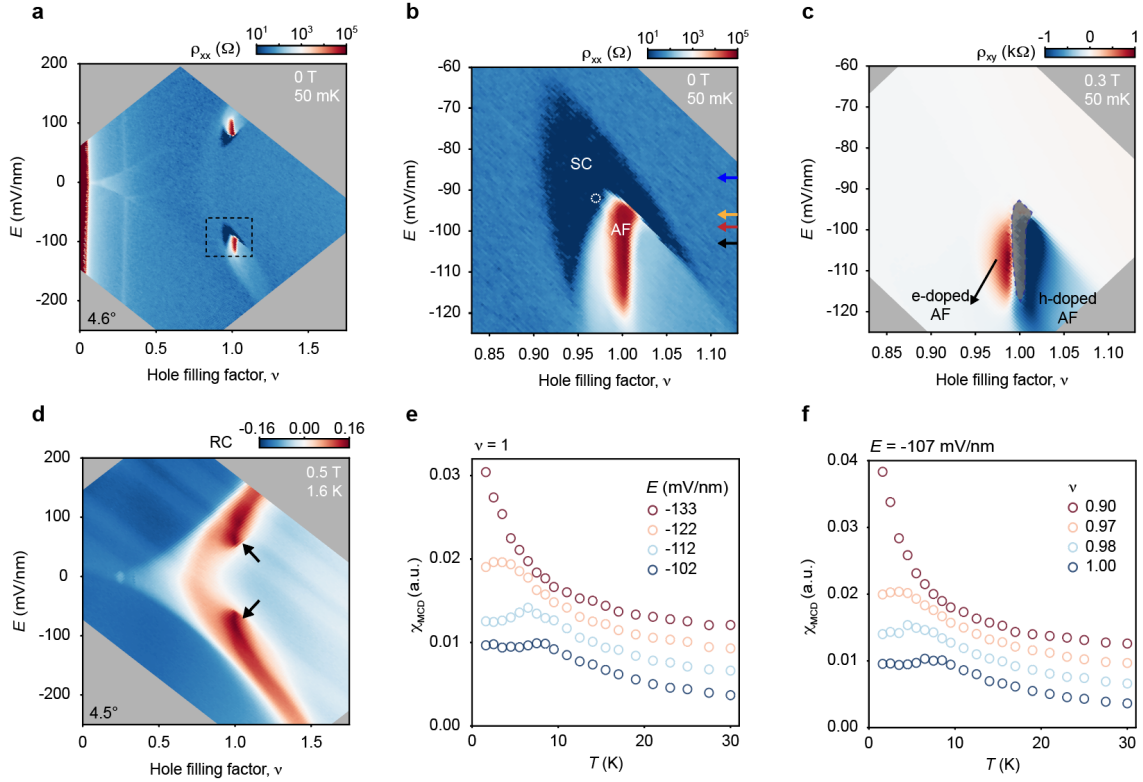

**Figure 2 | Superconductor and AF insulator. a,** Longitudinal resistivity $\rho_{xx}$ versus $\nu$ and $E$ at $T = 50$mK and $B = 0$T for device 1 (4.6° tWSe$_2$ transport device). **b,c,** Close-up view of the dashed box in **a** for $\rho_{xx}$ (**b**) and Hall resistivity $\rho_{xy}$ at $B = 0.3$T (**c**). Superconductor (SC), antiferromagnetic (AF) insulator, and electron-doped and hole-doped AF insulators are identified. The white circle in **b** marks the location with the highest $T_c$. Arrows denote the $E$-fields, under which the phase diagrams are mapped in Fig. 3. **d,** Reflection contrast (RC) at the moiré exciton resonance versus $\nu$ and $E$ at $T = 1.6$K and $B = 0.5$T for device 2 (4.5° tWSe$_2$ optics device). Arrows denote the AF insulator with enhanced RC. **e,f,** $T$-dependence of magnetic susceptibility $\chi_{MCD}$ at $\nu = 1$ and varying $E$-fields (**e**) and at $E = -107$mV/nm and varying filling factors (**f**). The curves are vertically displaced by 0.003 for clarity. $T_N$ decreases with more negative $E$-fields and with doping away from $\nu = 1$. In **a-d**, the grey regions are experimentally inaccessible.

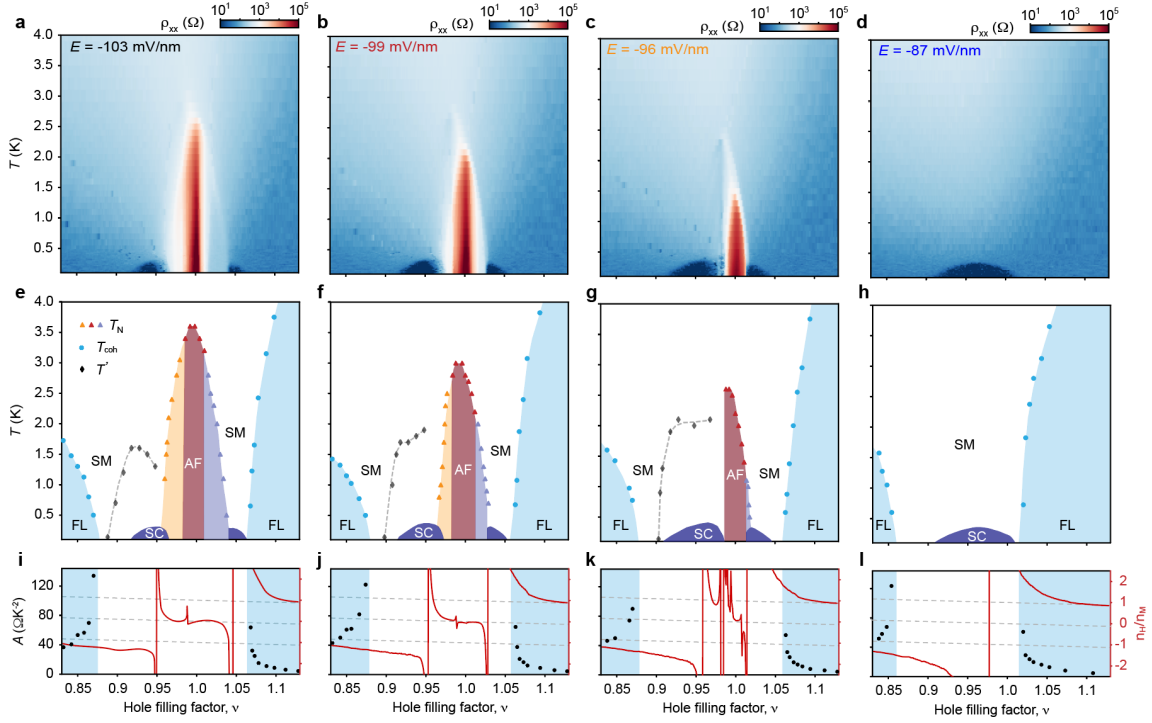

**Figure 3 | Phase diagrams. a-d,** $\rho_{xx}$ under $B = 0T$ versus $\nu$ and $T$. **e-h,** Extracted $\nu - T$ phase diagrams exhibiting antiferromagnetic insulators (AF, shaded in brown), electron-doped AF insulators (yellow), hole-doped AF insulators (purple), superconductors (SC, dark blue), strange metals (SM), and Fermi liquids (FL, light blue) with their corresponding temperature scales, $T_N$, $T_{coh}$, and $T'$ (symbols). **i-l,** Filling factor dependence of coefficient $A$ (symbols, left axis) and normalized Hall density $\frac{n_H}{n_M}$ (red lines, right axis) at $T = 50$mK. Coefficient $A$ was obtained by fitting $\rho_{xx} = \rho_0 + AT^2$ to experiment in the FL phase (shaded in light blue). Hall density was extracted from $\rho_{xy}$ measured under $B = 0.5T$. The dashed lines denote $\frac{n_H}{n_M} = 2 - \nu, 1 - \nu$, and $-\nu$. The four columns from left to right correspond to $E = -103$ mV/nm, $-99$ mV/nm, $-96$ mV/nm, and $-87$ mV/nm. Mott transition at $\nu = 1$ occurs at critical field $E_c \approx -92$mV/nm.

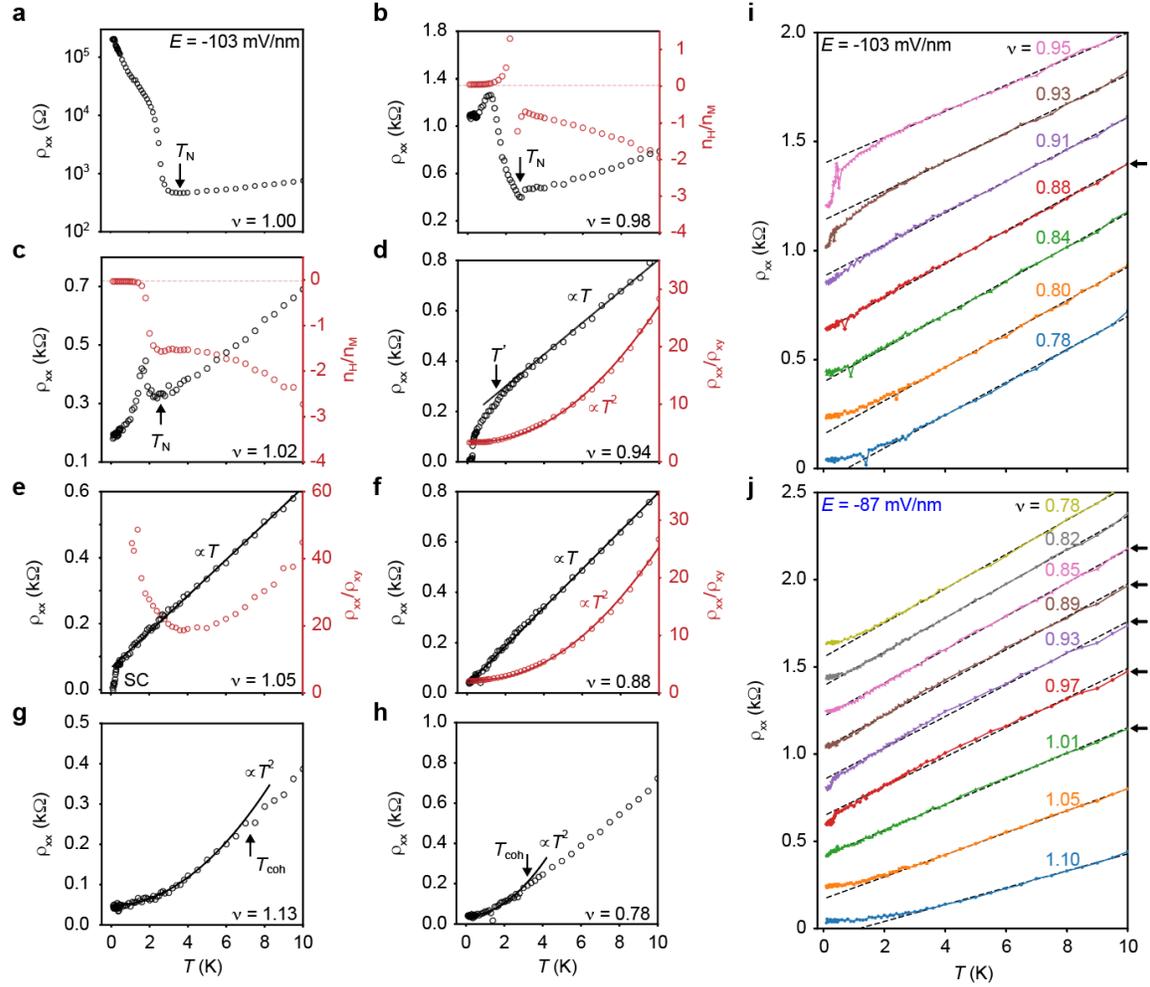

**Figure 4 | Temperature dependent transport at selected fillings and *E*-fields. a-h,** Temperature dependence of transport characteristics at $E = -103\,\text{mV/nm}$ for selected fillings: $\nu = 1$ (**a**), 0.98 (**b**), 1.02 (**c**), 0.94 (**d**), 1.05 (**e**), 0.88 (**f**), 1.13 (**g**), and 0.78 (**h**). Left axis: $\rho_{xx}$ under $B = 0\text{T}$ (black symbols for experiment, black lines for linear or quadratic fits); right axis: normalized Hall density $\frac{n_H}{n_M}$ (**b,c**) and Hall angle $\frac{\rho_{xx}}{\rho_{xy}}$ (**d-f**) determined under $B = 0.5\text{T}$ (red symbols for experiment, red lines for quadratic fits). Red dashed lines in **b,c** denote Hall density $(1 - \nu)$. Néel temperature $T_N$ is determined from the local resistivity minimum. Crossover temperature $T'$ is defined by 10% deviation from $T$-linear resistivity. Coherence temperature $T_{coh}$ is defined by 10% deviation from $T^2$ resistivity. **i,j,** $T$-dependence of $\rho_{xx}$ under $B = 0\text{T}$ at different filling factors for $E = -103\text{mV/nm}$ (**i**) and $-87\text{mV/nm}$ (**j**). The curves are vertically displaced by $0.2\,k\Omega$ for clarity. Black dashed lines are linear fits to the high-temperature resistivity. Arrows denote $T$-linear resistivity for the entire temperature range above $T_c$.

**Extended Data Figures**

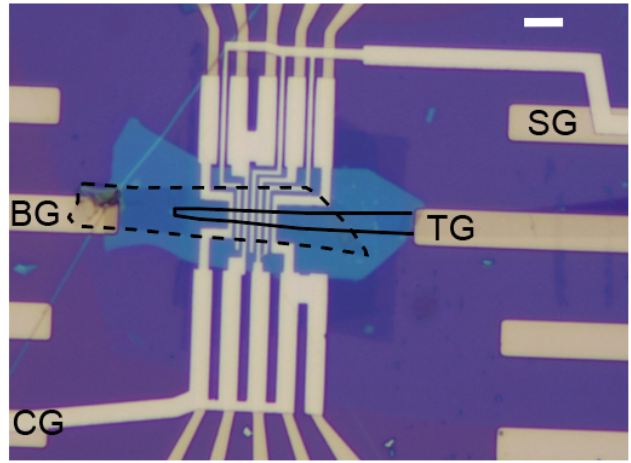

**Extended Data Figure 1 | Device optical image.** Optical micrograph of a 4.6º tWSe$_2$ device. The scale bar is $4\mu m$. The channel region is defined by the overlap area of the top gate (TG, solid line) and bottom gate (BG, dashed line). The contact gate (CG) and split gate (SG) are also labelled.

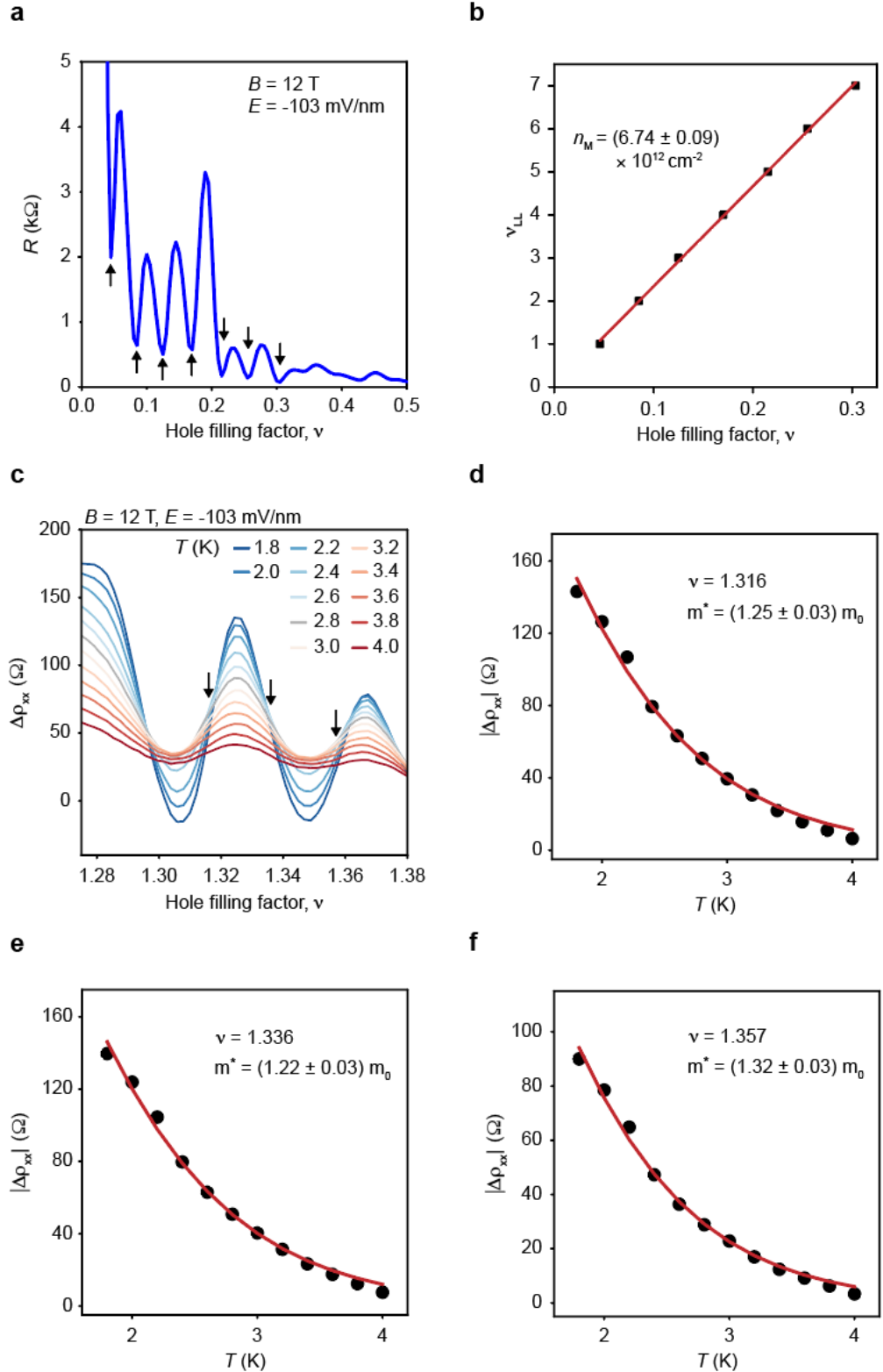

**Extended Data Figure 2 | Determination of the moiré density and the quasiparticle effective mass at $E = -103$mV/nm. a**, Longitudinal resistance $R$ as a function of $\nu$ at $B = 12$T, $T = 1.6$K, and $E = -103$mV/nm. The arrows label the positions of the Landau

levels with index $\nu_{LL} = 1-7$. **b,** Landau level index $\nu_{LL}$ as a function of $\nu$ that follows a linear dependence (red line). A moiré density $n_M \approx (6.74 \pm 0.09) \times 10^{12}\,\text{cm}^{-2}$ can be determined from the fitted slope. **c,** Resistivity difference $\Delta\rho_{xx} = \rho_{xx}(T) - \rho_{xx}(T = 4.4K)$ as a function of $\nu$ at varying $T$ and at $B = 12\text{T}$. No Shubnikov-de Haas oscillation can be observed at $T = 4.4\text{K}$. **d-f,** $T$-dependence of the amplitude $|\Delta\rho_{xx}|$ at $\nu = 1.316$ (**d**), 1.336 (**e**), and 1.357 (**f**) as marked by the arrows in **c**. $|\Delta\rho_{xx}|$ at each $\nu$ was obtained as the difference between the nearest $\Delta\rho_{xx}$ peak and dip centered around $\nu$. The quasiparticle effective mass $m^*$ was extracted by fitting the data to $|\Delta\rho_{xx}| = \frac{R_a \lambda(T)}{\sinh\lambda(T)}$ (red lines). Here $R_a$ is the amplitude and $\lambda(T) = \frac{2\pi^2 k_B T m^*}{\hbar eB}$.

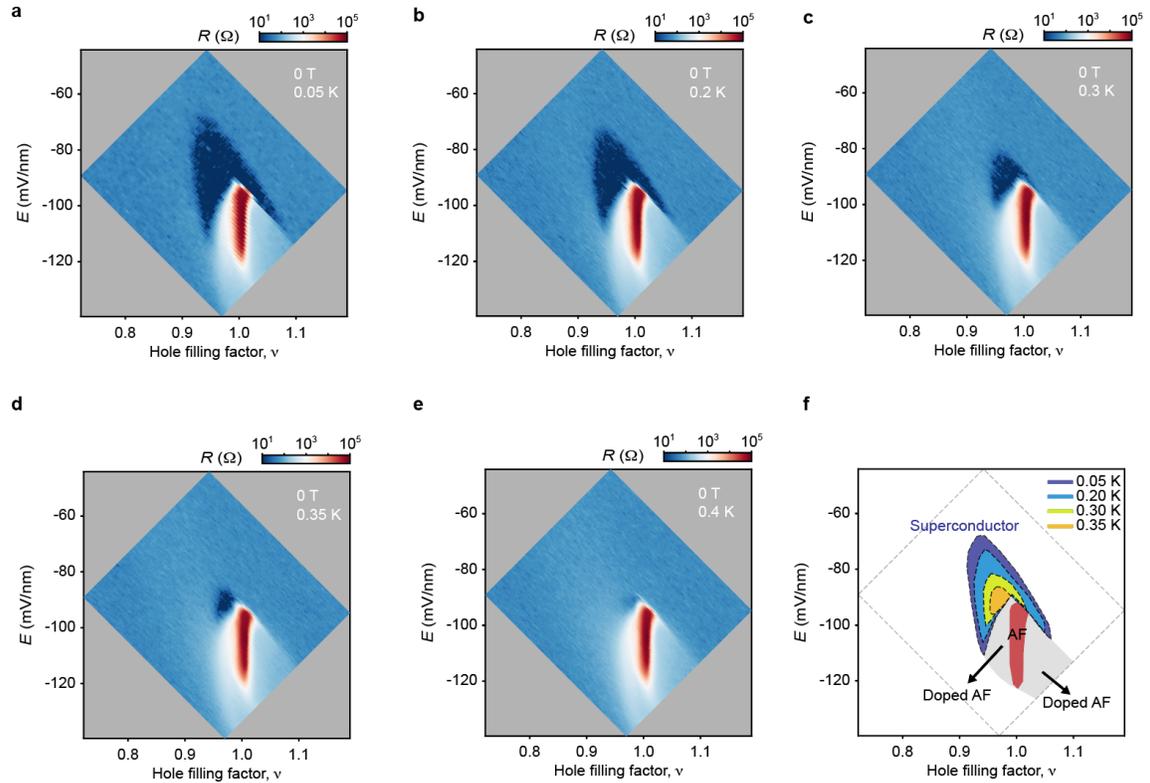

**Extended Data Figure 3 | Temperature dependent $\nu - E$ phase diagrams. a-e,** Longitudinal resistance $R$ as a function of $\nu$ and $E$ at $B = 0$T and $T = 0.05$K (**a**), 0.2K (**b**), 0.3K (**c**), 0.35K (**d**), and 0.4K(**e**). **f,** Phase diagram extracted from the data in **a-e**. The superconducting region shrinks with increasing temperature. The region immediately next to the tip of the AF insulator has the highest $T_c$.

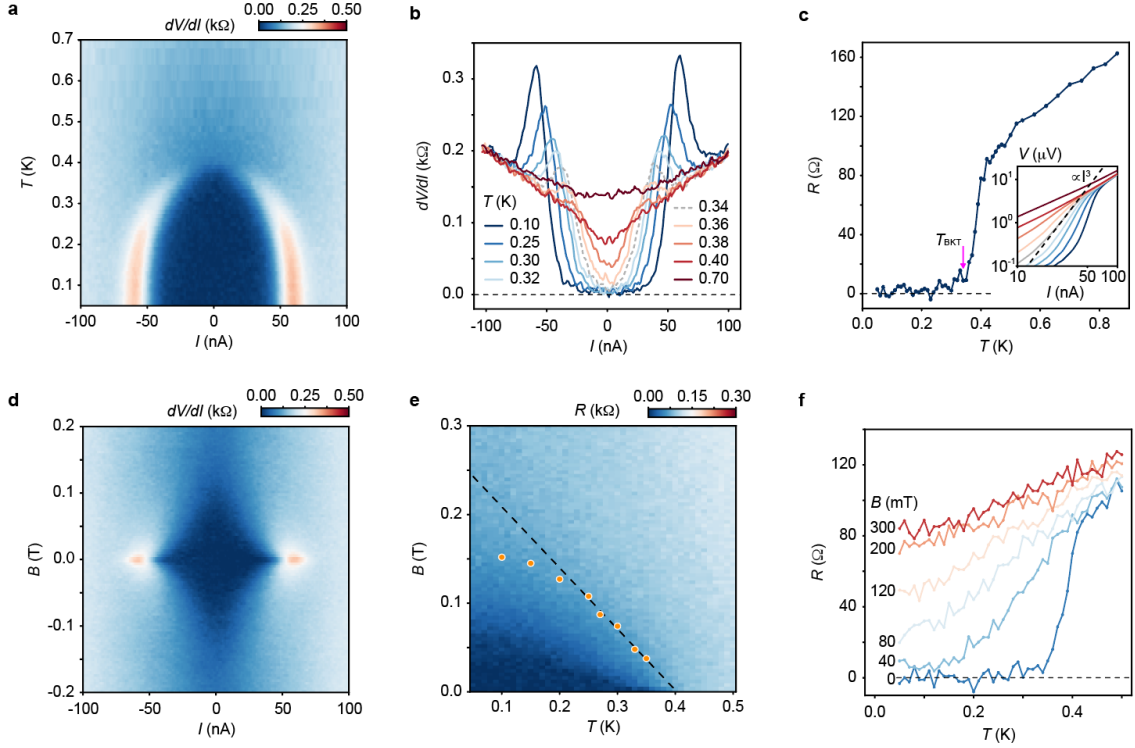

**Extended Data Figure 4 | Superconductivity at fixed $\nu \approx 0.97$ and $E \approx -92\,\text{mV/nm}$ with optimal $T_c$. a,** Differential resistance $\frac{dV}{dI}$ as a function of $T$ and DC bias current $I$ at $B = 0\,\text{T}$. **b,** Representative line cuts of **a** at different temperatures. Grey dashed line: $\frac{dV}{dI}$ at the Berezinskii-Kosterlitz-Thouless (BKT) transition temperature $T_{BKT} \approx 340\,\text{mK}$. **c,** $T$-dependence of zero-bias resistance $R$. The pink arrow labels $T_{BKT}$. Inset: The longitudinal voltage $V$ as a function of $I$. The line color is defined in **b**. At $T_{BKT}$, the $I-V$ dependence follows $V \propto I^3$ (dashed line). **d,** $\frac{dV}{dI}$ as a function of $B$ and $I$ at $T = 50\,\text{mK}$. **e,** $R$ as a function of $B$ and $T$ with the critical $B$-field $B_{c2}$ marked by the orange data points. The dashed line is a fit to $B_{c2} \approx \frac{\Phi_0}{2\pi\xi^2}\left(1 - \frac{T}{T_c}\right)$, from which a superconducting coherence length $\xi \approx 34\,\text{nm}$ is determined. **f,** Representative linecuts of **e** at varying $B$.

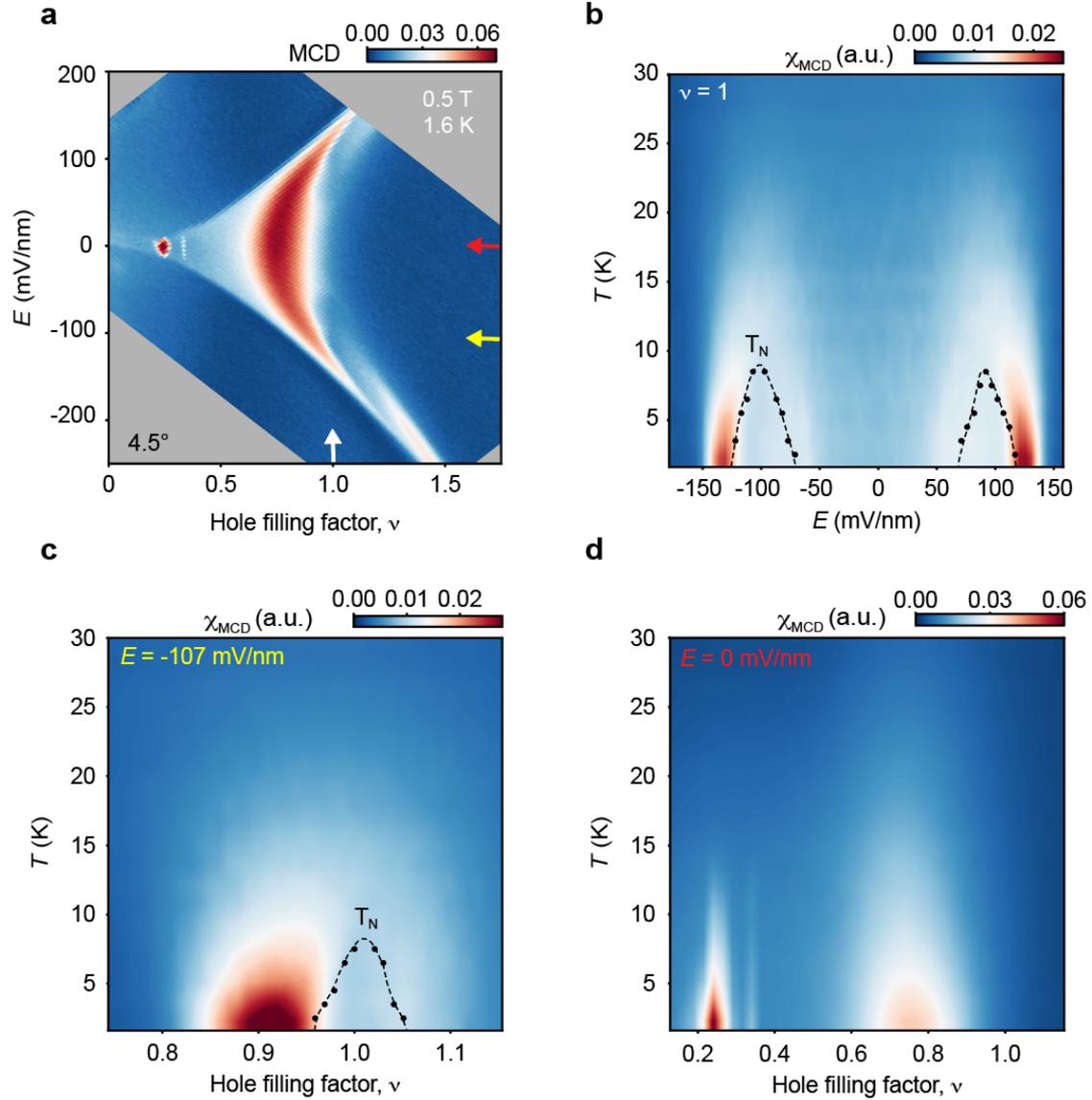

**Extended Data Figure 5 | MCD map and temperature dependence of the magnetic susceptibility. a,** MCD signal of a 4.5° tWSe$_2$ device versus $\nu$ and $E$ at $T = 1.6$K and $B = 0.5$T. A suppressed MCD is observed at the $\nu = 1$ insulating state. The arrows indicate the electric fields and filling factors used for the $T$-dependent studies in **b-d**. **b-d,** $\chi_{MCD}$ as a function of $E$ and $T$ at $\nu = 1$ (**b**), and that as a function of $\nu$ and $T$ at $E = -107$mV/nm (**c**) and 0mV/nm (**d**). $\chi_{MCD}$ is suppressed in regions bound by the black dashed lines in **b** and **c**, where AF order emerges. The data points show the extracted values of $T_N$. Evidence of AF ordering is observed only in the vicinity of $\nu = 1$ at finite $E$-fields. The enhanced $\chi_{MCD}$ near $\nu = 0.9$ in **c** and near $\nu = 0.8$ in **d** is correlated with the location of the vHS. The generalized Wigner crystals at $\nu = \frac{1}{4}, \frac{1}{3}$ in **d** also show enhanced $\chi_{MCD}$ at low temperatures because of local magnetic moments.

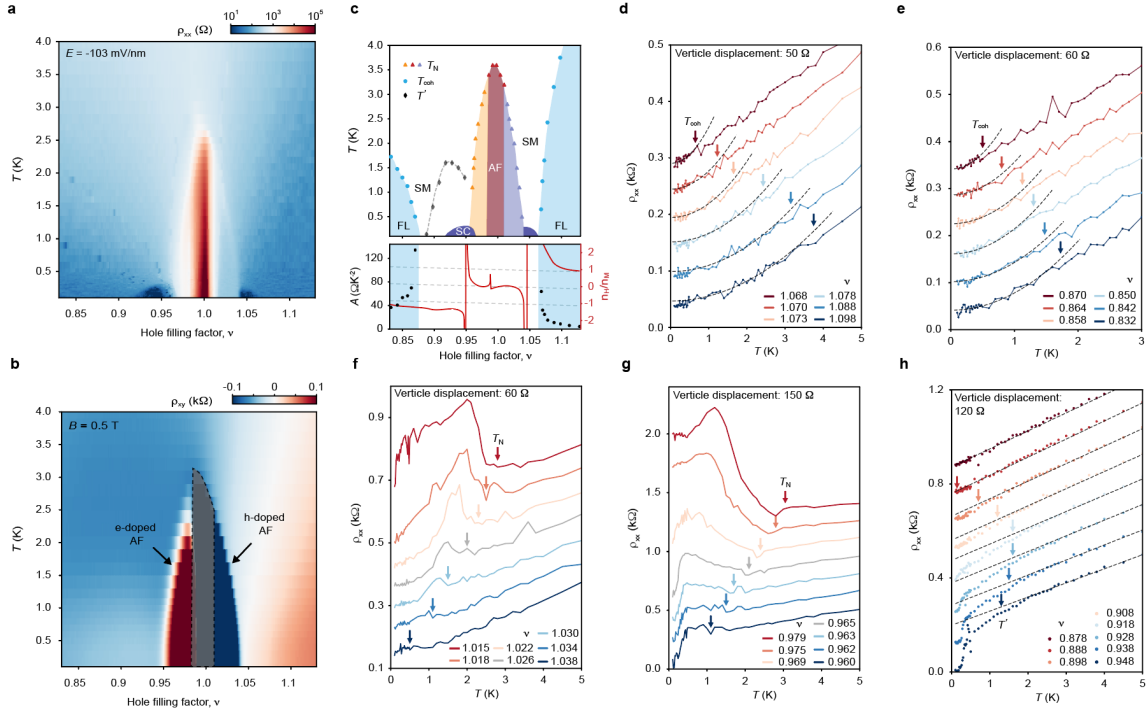

**Extended Data Figure 6 | Determination of temperature scales at $E = -103$mV/nm.**
**a,** $\rho_{xx}$ versus $\nu$ and $T$ at $B = 0$T (black arrow in Fig. 2b). **b,** $\rho_{xy}$ versus $\nu$ and $T$ measured at $B = 0.5$T. The grey-shaded region marks the AF insulator, where $\rho_{xy}$ cannot be reliably measured. The electron-doped and hole-doped AF insulators have low Hall densities; they develop only when the AF insulator is stabilized below $T_N \approx 3.5$ K. The Hall resistivity also changes sign at high temperatures which allows for identification of the vHS. **c,** Upper panel: Extracted $\nu - T$ phase diagrams corresponding to **a,b** (copy of Fig. 3e). Lower panel: Filling factor dependence of the coefficient $A$ and $\frac{n_H}{n_M}$ at $T = 50$mK (copy of Fig. 3i). **d-h**, $T$-dependence of $\rho_{xx}$ at representative fillings corresponding to the data points in **c** (blue dots for **d,e**; purple and orange triangles for **f** and **g**, respectively; black diamonds for **h**). The curves are vertically displaced for clarity. The black dashed lines in **d,e** are fits to the low-temperature part of the data by $\rho_{xx} = \rho_0 + AT^2$. We define $T_{coh}$ the temperature at which $\rho_{xx}$ deviates from the fit by 10% (arrows). $T_N$ in **f,g** is the temperature below which $\rho_{xx}$ starts to show a bump (arrows). The black dashed lines in **h** are linear fits to the high-temperature part of the data. We define $T'$ the temperature at which $\rho_{xx}$ deviates from the linear fit by 10% (arrows).

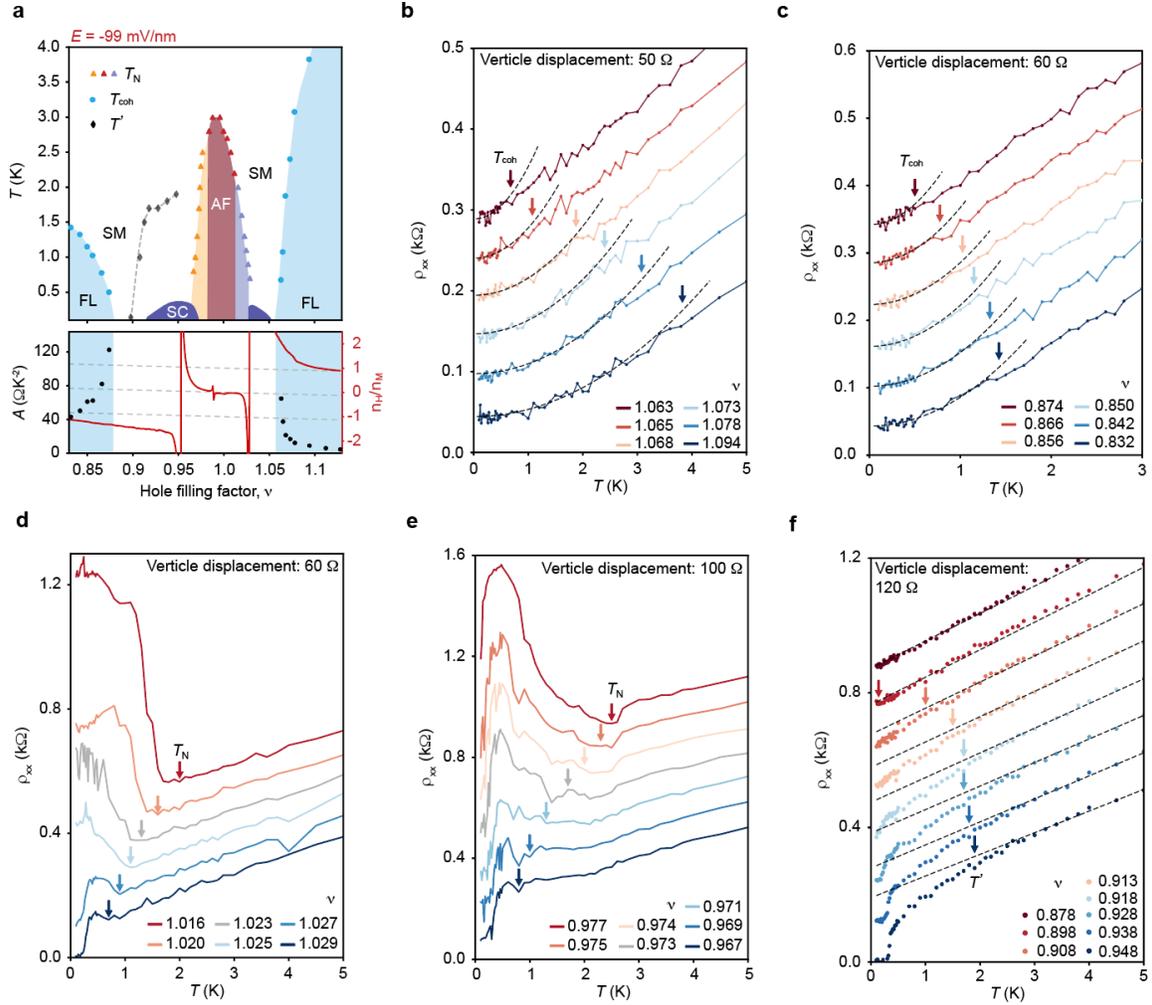

**Extended Data Figure 7 | Determination of temperature scales at $E = -99$mV/nm. a,** Copies of Fig. 3f,j in the main text. **b-f,** $T$-dependence of $\rho_{xx}$ at representative fillings corresponding to the data points in **a**. We follow the same notation and convention as in Extended Data Fig. 6.

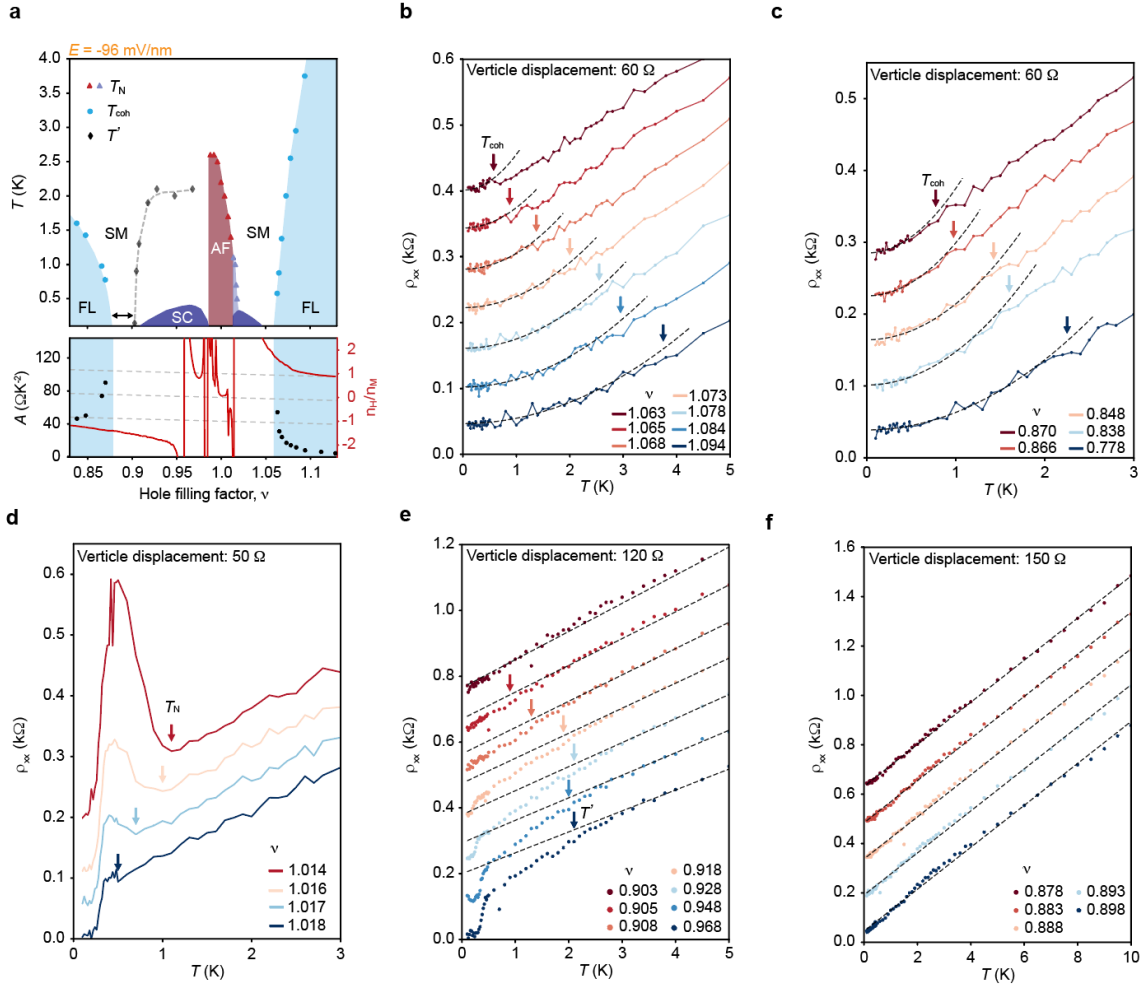

**Extended Data Figure 8 | Determination of temperature scales at $E = -96$ mV/nm. a,** Copies of Fig. 3g,k in the main text. **b-e,** $T$-dependence of $\rho_{xx}$ at representative fillings corresponding to the data points in **a** (blue dots for **b,c**; purple triangles for **d**; black diamonds for **e**). **f,** Linecuts in the strange metal region marked by the black arrow in **a**. The black dashed lines are linear fits to the data over the entire temperature range.

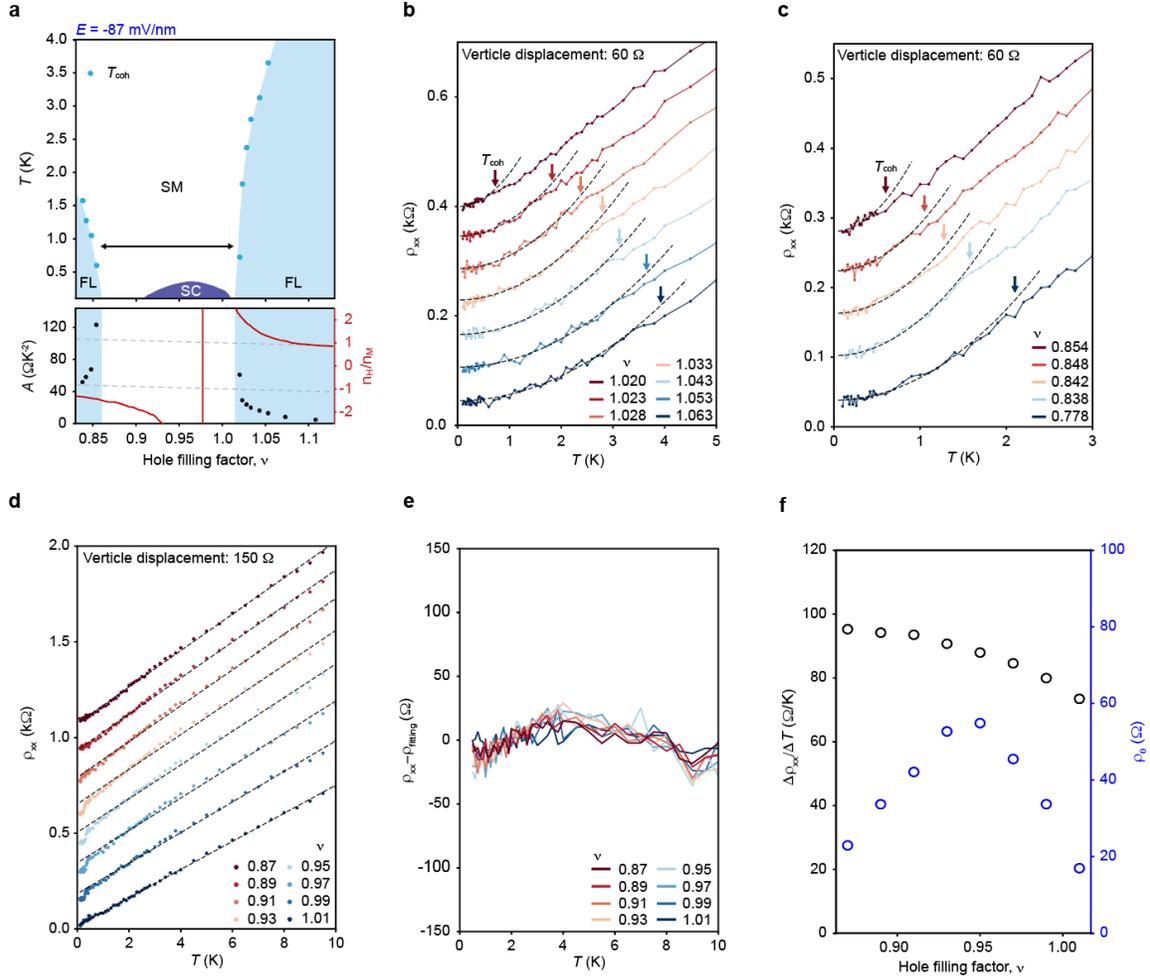

**Extended Data Figure 9 | Determination of temperature scales and analysis of the strange metal phase at $E = -87$ mV/nm. a,** Copies of Fig. 3h,l in the main text. **b,c,** $T$-dependence of $\rho_{xx}$ at representative fillings corresponding to the blue data points in **a**. **d,** Linecuts in the strange metal region marked by the black arrow in **a**. **e,** The deviation ($\rho_{xx} - \rho_{fitting}$) of the data from the linear fits in **d** plotted as a function of $T$. The small residuals demonstrate the near perfect $T$-linear dependence. **f,** Filling factor dependence of the slope $\Delta\rho/\Delta T$ (black) and the residual resistivity $\rho_0$ (blue) for the data in **d**. No clear correlation between $\Delta\rho/\Delta T$ and $\rho_0$ is observed.

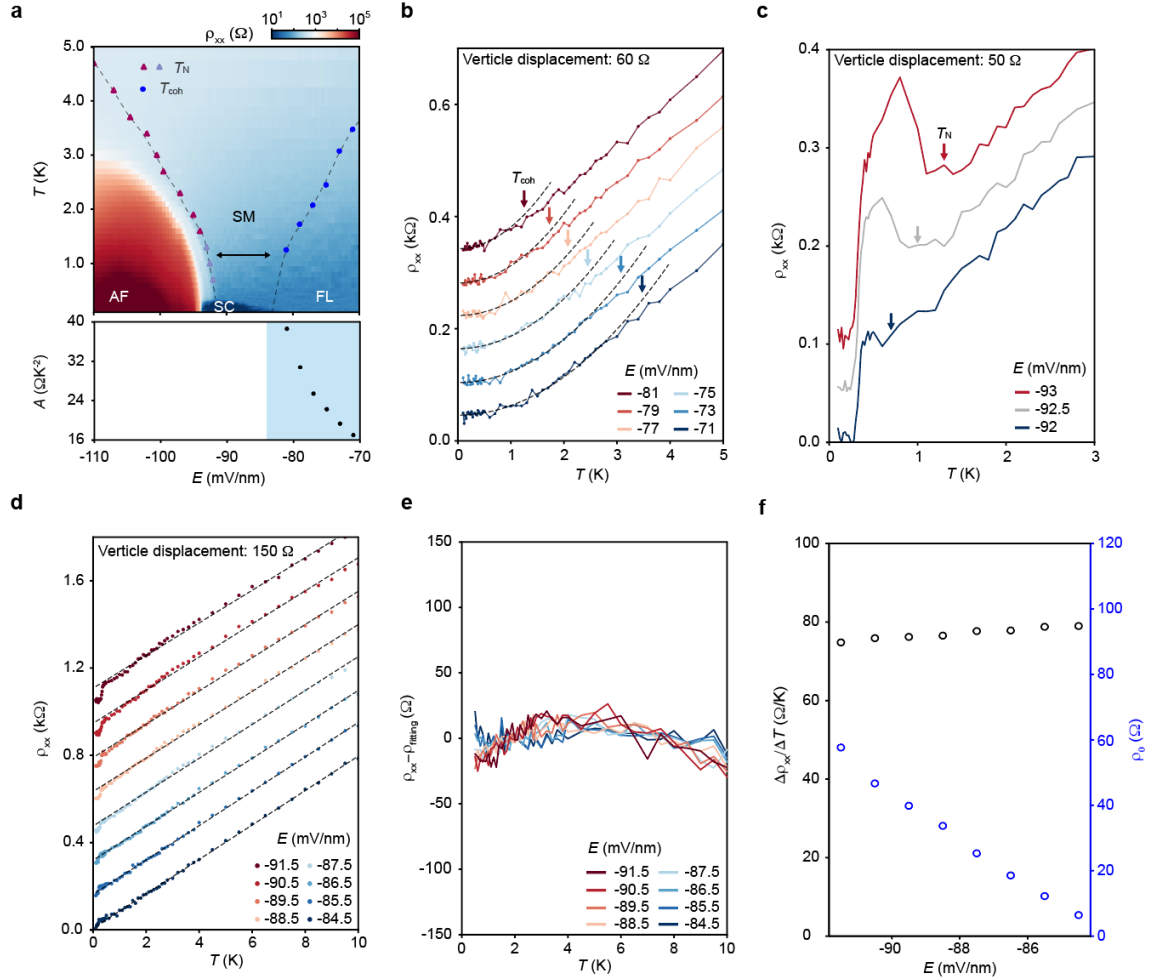

**Extended Data Figure 10 | Determination of temperature scales and analysis of the strange metal phase at $\nu = 1$. a,** Upper panel: $\rho_{xx}$ versus $E$ and $T$ at $B = 0$T and $\nu = 1$. The different phases together with their corresponding temperature scales ($T_N$ and $T_{coh}$) are labeled. The arrow denotes the range for the strange metal phase. Lower panel: the dependence of the coefficient $A$ on $E$-field in the Fermi liquid phase. As $E$-field approaches the Mott transition from the insulator side, $T_N$ decreases continuously and vanishes at $E_c$, beyond which a superconducting dome immediately appears. Moreover, a clear AF metallic state over the temperature range $T_C < T < T_N$ is observed right next to the AF insulator near $E_c$. **b,c,** $T$-dependence of $\rho_{xx}$ at representative $E$-fields corresponding to the blue data points (**b**) and the grey triangles (**c**) in **a**. Two thermodynamic phase transitions are observed in **c**: a normal metal to an AF metal at $T_N$ and an AF metal to a superconductor at $T_c$. **d,** Linecuts in the strange metal region marked by the black arrow in **a**. **e,** The deviation ($\rho_{xx} - \rho_{fitting}$) of the data from the linear fits in **d** plotted as a function of $T$. **f,** $E$-field dependence of $\Delta\rho/\Delta T$ (black) and $\rho_0$ (blue) for the data in **d**. No clear correlation between $\Delta\rho/\Delta T$ and $\rho_0$ is observed. The results suggest a strange metal phase above the superconducting dome over an extended range of $E$-fields.

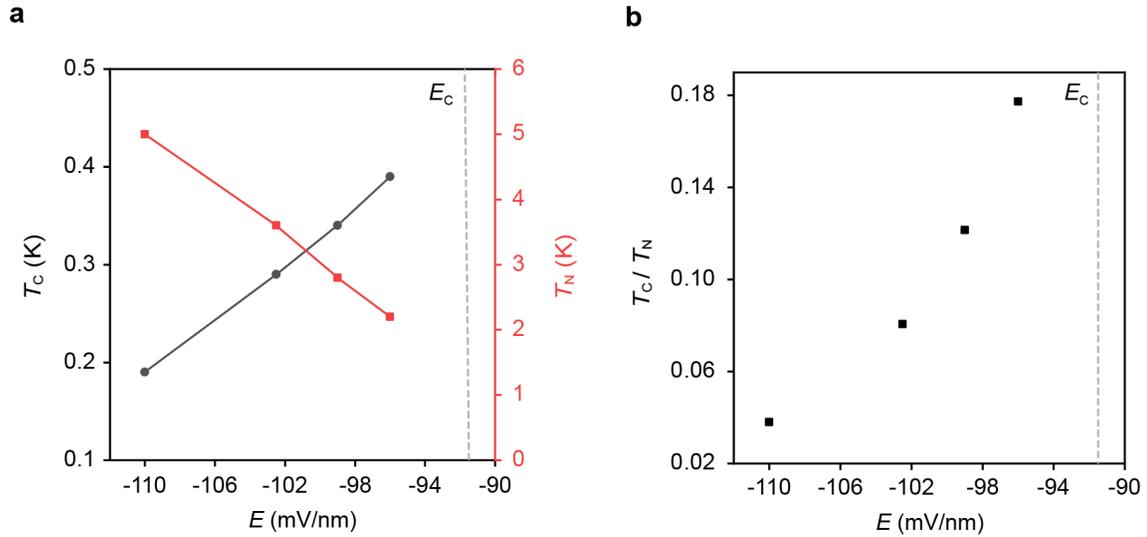

**Extended Data Figure 11 | Electric field dependence of $T_c/T_N$ a,** Optimal $T_c$ (black) and $T_N$ for the AF insulator (red) as a function of $E$-field at $B = 0T$. The optimal $T_c$ at a given $E$-field is the highest $T_c$ from the $\nu - T$ phase diagram in Fig. 3. The grey dashed line labels the critical $E$-field $E_c$ at the Mott transition. **b,** $T_c/T_N$ as a function of $E$-field; the ratio increases substantially near $E_c$.